\documentclass[twoside,11pt]{article}
\usepackage{graphicx}
\usepackage{bm}
\usepackage{float}
\usepackage{appendix}
\usepackage{tabularray}
\usepackage{booktabs}
\usepackage{lipsum}
\usepackage{cite}
\usepackage{appendix}
\usepackage[margin=1in]{geometry}
\usepackage[usenames]{color}
\usepackage{amsfonts, amssymb, amsmath, amsthm, multicol}
\usepackage[colorlinks=true, pdfstartview=FitV, linkcolor=blue,
citecolor=blue, urlcolor=blue]{hyperref}
\usepackage[usenames]{color}
\definecolor{Red}{rgb}{1,0.05,0}
\definecolor{Grn}{rgb}{0.1,0.7,0.1}
\definecolor{Blu}{rgb}{0.1,0.1,0.6}
\definecolor{Org}{rgb}{1,0.45,0}
\definecolor{Vio}{rgb}{0.6578,0,0.9478}
\definecolor{Mag}{rgb}{1,0.2,0.3}
\usepackage{authblk}
\usepackage{booktabs}
\usepackage{breqn}

\usepackage[section]{placeins}
\usepackage{graphicx}
\usepackage{amssymb}
\usepackage{epstopdf, epsfig}
\usepackage{color}
\usepackage{float}
\usepackage{amsmath}
\usepackage{tabularx}
\usepackage{bm}
\usepackage{accents}
\usepackage[numbers]{natbib}
\usepackage{array}
\usepackage{tabularx}
\usepackage{multirow,multicol, makecell}
\usepackage{pdflscape}
\usepackage{float}
\usepackage{breqn}
\usepackage{mathtools}
\usepackage{breqn}
\usepackage[flushleft]{threeparttable}
\usepackage{booktabs,caption}
\usepackage{footnote}
\usepackage{rotating}
\usepackage{lscape}
\makesavenoteenv{tabular}
\newcolumntype{C}[1]{>{\centering\arraybackslash}p{#1}}
\newcolumntype{L}[1]{>{\raggedright\arraybackslash}p{#1}}
\newcolumntype{M}[1]{>{\centering\arraybackslash}m{#1}}
\usepackage[mathlines,displaymath]{lineno}
\usepackage{lscape}
\usepackage{siunitx}
\usepackage{mathtools}
\usepackage[labelfont=bf]{caption}
\usepackage{hyperref}
\hypersetup{
    colorlinks,
    linkcolor={red!50!black},
    citecolor={blue!50!black},
    urlcolor={blue!80!black}
}

\usepackage{adjustbox}
\usepackage{xcolor}
\usepackage{soul}

\numberwithin{rmk}{section}
\numberwithin{nt}{section}

\title{Droplet at the Corner of a V-Shaped Fiber}

\author[1]{\textcolor{black}{Yi Zhang}}
\author[1]{\textcolor{black}{Apurav Tambe}}
\author[1]{\textcolor{black}{Zhao Pan}\thanks{To whom correspondence may be addressed: {zhao.pan}@uwaterloo.ca}}
\affil[1]{University of Waterloo, Department of Mechanical and Mechatronics Engineering, Waterloo, ON, Canada}

\date{\today}

\begin{document}
\maketitle \vspace{1cm}
% \linenumbers
\section*{Abstract}
\textit{Hypothesis}

\noindent A fundamental question in the physics of droplet--fiber interactions is: What is the maximum droplet volume a fiber can retain?
While this problem has been studied for horizontal fibers and at the apex $\Lambda$-shaped bent fibers, it remains less explored for V-shaped bent fibers, despite their demonstrated advantages in engineering applications such as fog harvesting. 
This work investigates the capability of V-shaped fibers in retaining droplets against gravity.
We hypothesize that the maximum droplet volume at the corner of a V-shaped fiber is influenced by the opening angle ($\alpha$) of the V-shape, and an optimal $\alpha$ may exist.

\noindent \textit{Methodologies}

\noindent An analytical model to predict the maximum droplet volume on V-shaped fibers is developed based on free energy analysis, and validated against experimental data from five liquid--fiber pairs.

\noindent \textit{Findings}

\noindent The dependence of the maximum droplet volume on $\alpha$ can be reasonably captured by the function $\cos\beta/\cos\left(\beta-\alpha/2\right)$, where $\beta$ denotes the droplet's off-axis angle.
As $\alpha$ increases from $0^\circ$ to $180^\circ$, the maximum droplet volume slightly decreases before entering a broad transition region around $\alpha \approx 40^\circ \text{--} 100^\circ$, and then increases at larger $\alpha$.
%\zpr{these two sentences are too detailed, boring, and hard to follow. Say something else, like what you achieved or something you can brag about.} meanwhile, the droplet's off-axis angle ($\beta$) increases continuously. This non-monotonic behavior of the maximum volume is primarily governed by the term $\cos\beta/\cos\left(\beta-\alpha/2\right)$.

\noindent \textbf{Key words}: Droplet; V-shaped fiber; Maximum volume; Free energy; Off-axis angle

\section{Introduction}

Droplet--fiber interactions are prevalent in both natural systems and engineering applications involving fibrous media, such as water harvesting~\cite{chen2022transport, park2013optimal}, filtration and separation~\cite{contal2004clogging, wurster2015bubbling}, functional textiles~\cite{zhang2023facile, miao2021biomimetic}, and microfluidic devices~\cite{gilet2009digital, khattak2024directed}. 
The physics underlying these interactions has been extensively studied. 
Dynamic behaviors of droplets are explored in various configurations, including droplets traveling along a helical fiber \cite{darbois2015droplets}, a tilted fiber \cite{gilet2010droplets}, a conical fiber \cite{van2021capillary}, and a vertical fiber or fiber array \cite{leonard2023droplets}. 
The capture mechanisms \cite{safavi2021droplet} and dispersion behaviors \cite{huang2024dispersion} of droplets impacting a fiber have also been investigated. 
On the other hand, studies on static cases have focused on the conformation and stability of droplets on a horizontal fiber \cite{carroll1986equilibrium, chou2011equilibrium}, a tilted fiber \cite{huang2009equilibrium}, and a fiber hub \cite{zhang2025droplets}.

\begin{figure}[!h]
    \centering\includegraphics[width=0.6\linewidth]{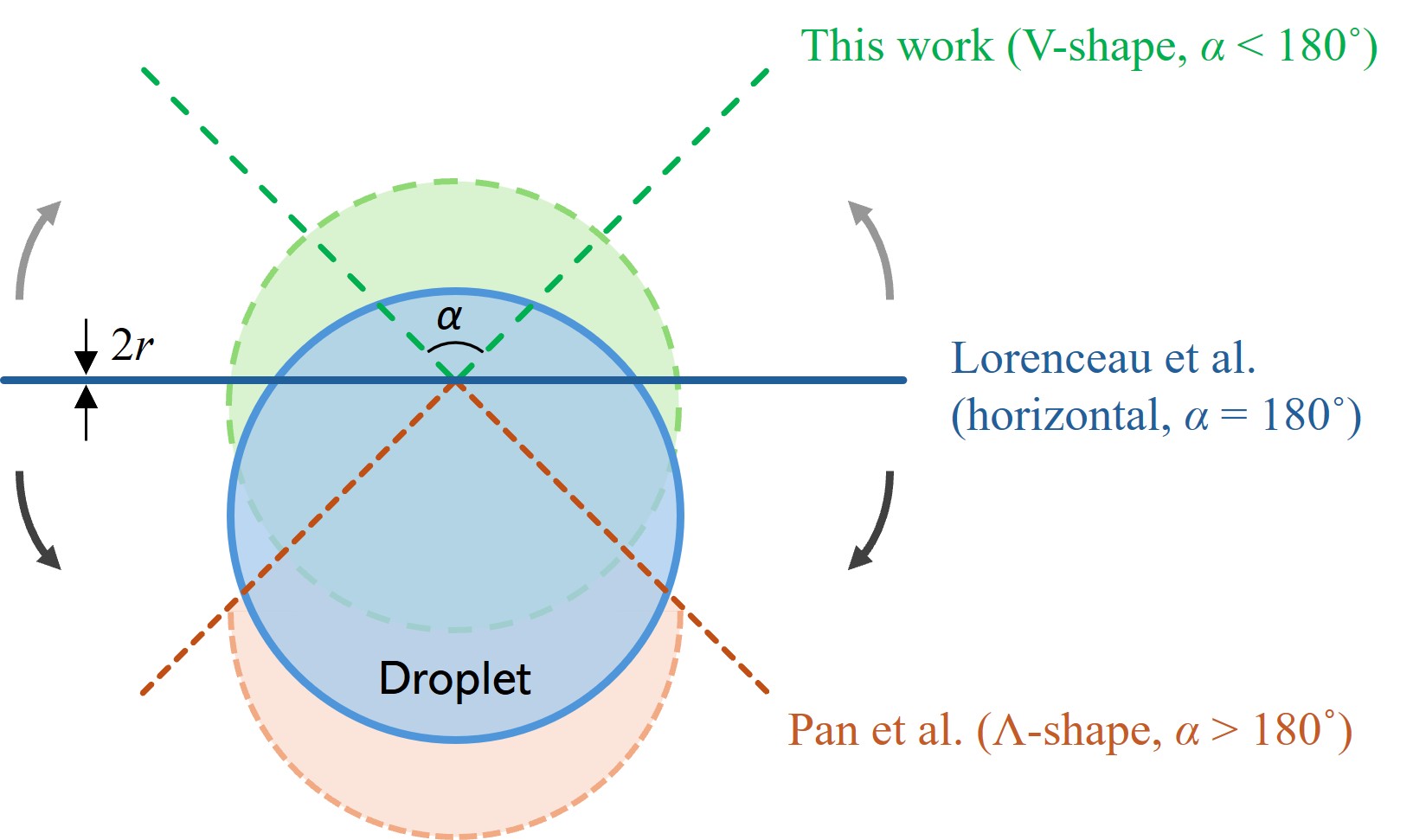}
    \caption{Droplet on a horizontal fiber ($\alpha=180^\circ$), a $\Lambda$-shaped fiber ($\alpha>180^\circ$), and a V-shaped fiber ($\alpha<180^\circ$). $\alpha$ is the opening angle of a fiber, and $r$ is the fiber radius.
    %\zpl{some ideas about this figure. You have 3 different setups, you can use different color groups to differentiate them, for horizontal shades of blue, Pan et al 2018, shades of red or purple etc. also put V-shape, and $\Lambda$-shape in the figure. Then people don't have to read the caption.}
    }
    \label{fig:literature review}
\end{figure}

Within the broad scope of existing research, a fundamental question concerns the maximum volume of a droplet that can be retained on a fiber. 
Lorenceau et al. \cite{lorenceau2004capturing} investigated this problem for a horizontal fiber, which can be thought of as a bend fiber with an opening angle ($\alpha$) being $180^\circ$, 
%corresponding to the case of $\alpha=180^\circ$, where $\alpha$ denotes the opening angle of a fiber, 
as illustrated in Figure~\ref{fig:literature review}. 
They derived a simple expression for the maximum droplet volume, given by $4\pi r\lambda^2$, where $r$ and $\lambda$ are the fiber radius and the capillary length of the liquid, respectively.
Pan et al. \cite{pan2018drop} extended the study to $\Lambda$-shaped bent fibers with $\alpha>180^\circ$, and found that the maximum droplet volume varies with $\alpha$, reaching a peak at $\alpha \approx 324^\circ$.\footnote{Note that the definition of the opening angle in this work differs from that in \cite{pan2018drop}.} 
%\zpl{add a footnote clarifying the definition of the angle in this work and Pan 2016 are different.}
However, the same question for V-shaped bent fibers with $\alpha < 180^\circ$ has been rarely addressed in the literature, despite demonstrated engineering applications.
Notably, V-shaped fibers have been employed in fog harvesting, where they enhance the efficient removal of water droplets and help prevent mesh clogging \cite{li2019fog, kennedy2024bio}.

Therefore, the objective of this paper is to investigate how V-shaped fibers retain liquid droplets against gravity.
Accordingly, section~\ref{sec: methods} describes the experimental methods used to characterize the maximum droplet volume of different liquids at the corner of V-shaped fibers with varying opening angles $\alpha$.
Section~\ref{sec: Results} presents the experimental results and develops theoretical models for predicting the maximum droplet volume. 
Last, the established models are validated against experimental data, and the study is concluded in section~\ref{sec: Conclusion}.

\section{Methods}
\label{sec: methods}

\begin{figure}[!h]
     \centering
     \includegraphics[width=0.7\linewidth]{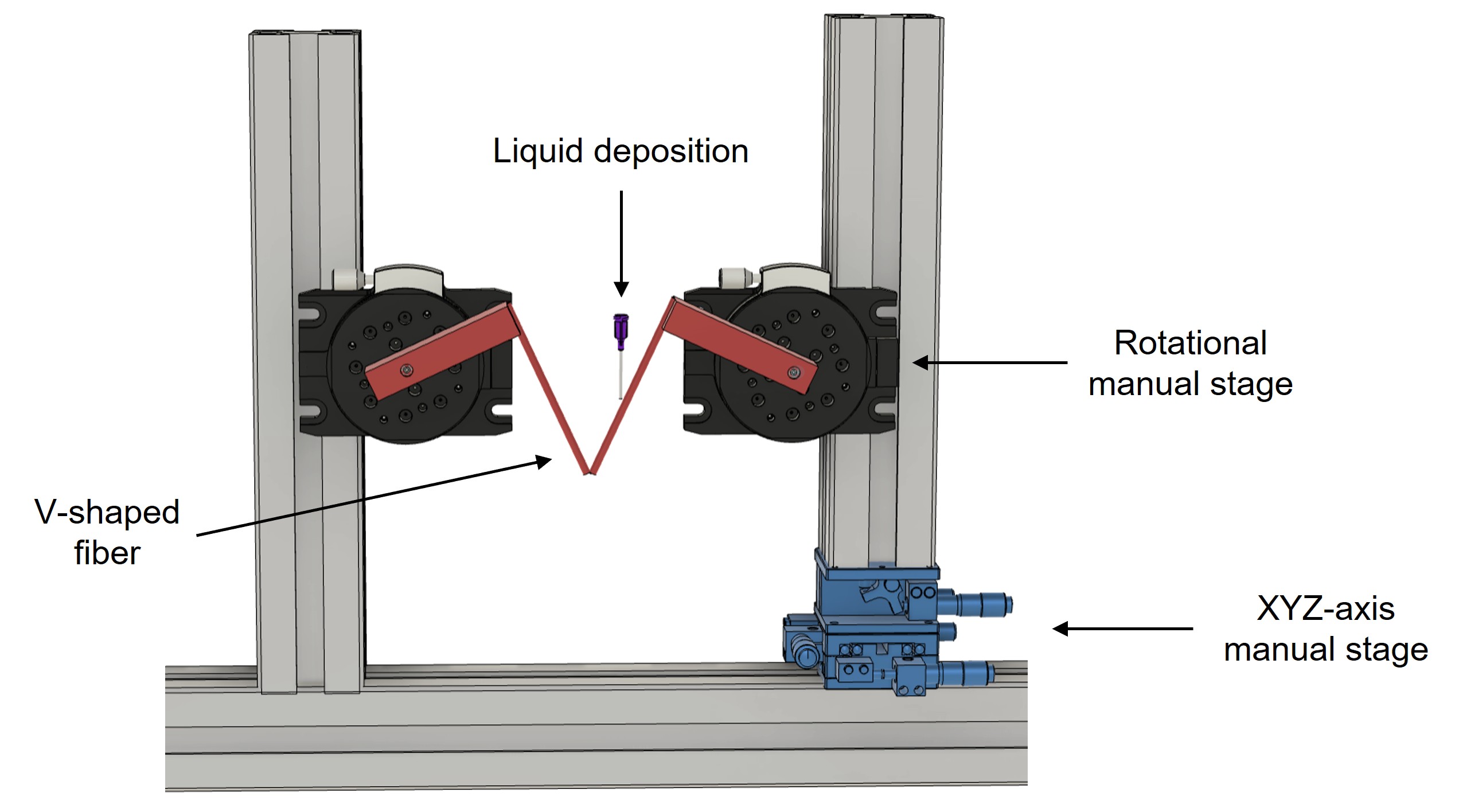}
     \caption{Experiment setup for measuring the maximum volume of droplets suspended at the corner of a V-shaped fiber. The positioning resolutions of the XYZ-axis stage and the rotational stage are 0.01~mm and 2~degrees, respectively.}
     \label{fig:exp setup}
\end{figure}

Figure~\ref{fig:exp setup} illustrates the experiment setup used to measured the maximum volume of droplets formed at the corner of a V-shaped fiber.
The V-shaped fiber is constructed by symmetrically mounting two individual fibers onto rotational manual stages, which allow precise adjustment of the opening angle~$\alpha$.
An XYZ-axis translation stage is employed to align and connect the two fibers at their ends. 
Fibers used in experiments include 0.50~mm-diameter glass fibers, 0.33~mm-diameter polyacrylic acid (PAA) coated copper fibers, as well as 0.20~mm- and 0.50~mm-diameter stainless steel fibers.

Three liquids, including silicone oil, \SI{0.01}{M} sodium dodecyl sulfate (SDS) solution, and deionized water, are employed. 
Measurements of the maximum droplet volume are performed for the following solid–-liquid pairs: water on glass fibers, silicone oil on stainless steel fibers and PAA-coated fibers, and SDS solution on PAA-coated fibers. 
All test combinations exhibit good wettability, characterized by small intrinsic contact angles ($\theta$), i.e., $\cos\theta\approx1$.
The surface tensions ($\gamma$) of the three liquids are \SI{21}{mN/m} for silicone oil, \SI{34}{mN/m} for \SI{0.01}{M} SDS solution, and \SI{72}{mN/m} for water. 
The densities ($\rho$) are \SI{0.93e3}{kg/\cubic\meter} for silicone oil and \SI{1.00e3}{kg/\cubic\meter} for both \SI{0.01}{M} SDS solution and water. 
%(can we add a table here? first column liquid and other columns with properties and fibers with which experiment performed. )

Prior to experiments, the fibers are cleaned by sequential rinsing with ethanol and deionized water, followed by air drying. 
The test liquid is loaded into a micropipette (Eppendorf, Germany) and carefully deposited onto one side of the V-shaped fiber. 
The deposited liquid travels along the side fiber toward the corner of the V-shape, where it accumulates and gradually forms a droplet through successive additions. 
As the maximum volume is approached, the droplet volume is increased in an increment of \SI{0.1}{\micro\liter} until detachment occurs. 
The profile of the droplet at its maximum volume is recorded using a digital camera (Nikon 750D, Japan) equipped with a \SI{105}{mm} macro lens.

\section{Results and discussion}
\label{sec: Results}

\subsection{Experimental results}
\label{sec: Exp results}

\begin{figure}[!h]
    \centering\includegraphics[width=1\linewidth]{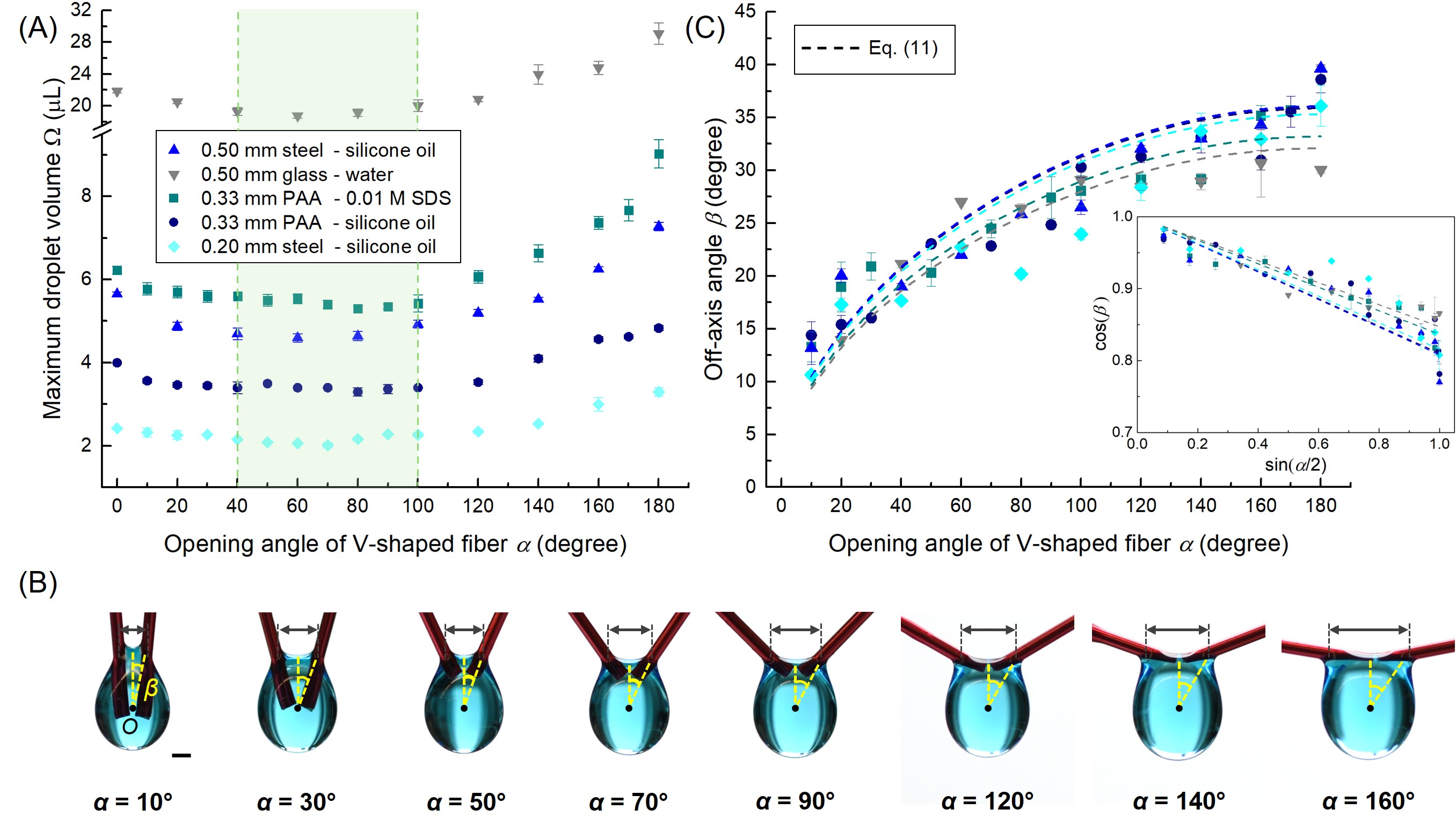}
    \caption{(A) Measured maximum droplet volumes at the corner of V-shaped fibers as a function of opening angle $\alpha$ (ranging from $0^\circ$ to $180^\circ$). Error bars represent the standard deviation from 3-5 repeated measurements. 
    (B) Photos of maximum-volume droplets at the fiber corner for $\alpha=10^\circ, 30^\circ, 50^\circ, 70^\circ, 90^\circ, 120^\circ, 140^\circ$, and $160^\circ$ (\SI{0.01}{M} SDS solution on \SI{0.33}{mm} PAA coated fibers). The off-axis angle $\beta$ is marked by dashed yellow lines. Scale bar indicates \SI{0.5}{mm}. 
    (C) Variation of $\beta$ with $\alpha$ (point symbols correspond to those in (A)).}
    \label{fig:exp data}
\end{figure}

Figure~\ref{fig:exp data}(A) presents the measured maximum volume ($\Omega$) of droplets at the corner of V-shaped fibers with opening angle ($\alpha$) varying from $0^\circ$ to $180^\circ$. 
Five liquid--fiber combinations are tested: silicone oil on \SI{0.5}{mm} stainless steel fibers (upward triangles), deionized water on \SI{0.5}{mm} glass fibers (downward triangles), \SI{0.01}{M} SDS solution on \SI{0.33}{mm} PAA coated fibers (squares), silicone oil on \SI{0.33}{mm} PAA-coated fibers (filed circles), and silicone oil on \SI{0.2}{mm} stainless steel fibers (diamond symbols). 
Across all tested cases, $\Omega$ exhibits a mild non-monotonic trend with respect to $\alpha$: it slightly decreases with increasing $\alpha$ before entering a broad transition region around $\alpha \approx 40^\circ \text{--} 100^\circ$ (highlighted by the green-shaded region); and then $\Omega$ increases for larger $\alpha$.

Figure~\ref{fig:exp data}(B) shows the morphology of maximum-volume droplets for varying $\alpha$.
The droplet's center of mass (COM), labeled as point $O$, is estimated by approximating the droplet shape as a sphere (see electronic support information (ESI) section~1).
For visual comparison, all droplet profiles are aligned such that their COMs lie on the same horizontal level.
As $\alpha$ increases, the spreading of droplet in the horizontal direction is enhanced (highlighted by the arrowed line segments), leading to an increment of the off-axis angle $\beta$, which is the angle between the vertical and the line connecting COM to the three-phase contact point where the fiber exits the droplet.
As will be discussed in Section \ref{sec: modeling}, 
% \zpl{where? cross ref here}
$\beta$ is a key parameter in determining the movement of three-phase contact points when the droplet's COM is perturbed.
Figure~\ref{fig:exp data}(C) shows the variation of $\beta$ with increasing~$\alpha$, revealing a clear upward trend.

\subsection{Modeling}
\label{sec: modeling}

Based on the experimental data and observations presented in Section \ref{sec: Exp results}, theoretical models for the maximum volume of droplets suspended at the corner of a V-shaped fiber are established based on a free energy analysis similar to that in \cite{pan2018drop}. 

\begin{figure}[!h]
    \centering\includegraphics[width=0.9\linewidth]{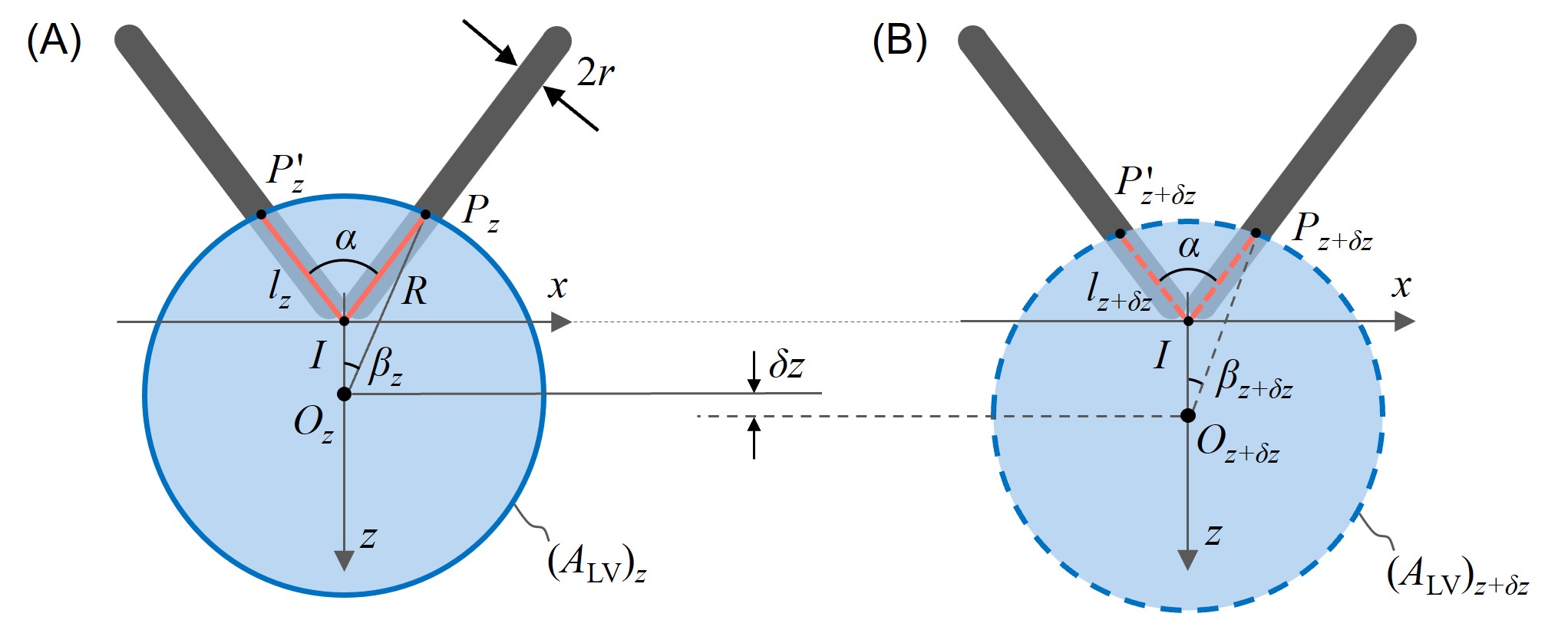}
    \caption{Schematic of a droplet hanging at the corner of a V-shaped fiber: (A) before and (B) after an infinitesimal perturbation of distance $\delta z$. The solid-liquid contact angle is not drawn. Subscripts $z$ and $z+\delta z$ denote the states before and after the perturbation, respectively. $A_{LV}$ is the liquid-vapor interfacial area; $I$ is the intersection point of fibers; $O$ is the droplet's center of mass; $P$ and $P'$ are the three-phase contact points; $R$ is the droplet radius; $l$ is the wetting length along one fiber; $r$ is the fiber radius; $\alpha$ is the opening angle of the V-shaped fiber; $\beta$ is the off-axis angle.}
    \label{fig:schematic}
\end{figure}

Figure~\ref{fig:schematic}(A) shows the schematic of a droplet suspending at the corner of a V-shaped fiber with a fiber diameter of $2r$ and an opening angle of $\alpha$. 
The droplet with a volume of $\Omega$ is assumed to be spherical \cite{lorenceau2004capturing, pan2018drop} with a radius of $R$ and an off-axis angle of $\beta$. 
The total wetting length of the fiber is $2l$. As gravity is the only body force, the total free energy ($G$) can be expressed as 
\begin{equation}
\label{eq:G}
G = \gamma_{LV} A_{LV} + \gamma_{SV} A_{SV} + \gamma_{LS} A_{LS} - \rho g \Omega z,
\end{equation}
where $\gamma$ and $A$ denote interfacial energy and area, respectively; 
subscripts $LV$, $SV$, and $LS$ represent liquid-vapor, solid-vapor, and liquid-solid interfaces, respectively. 
$\rho$ and $g$ are the density of liquid and gravitational constant, respectively;
$z$ describes the height of the droplet's center of mass.
Assuming that Young's equation ($\gamma_{LV}\cos\theta + \gamma_{LS} = \gamma_{SV}$, where $\theta$ is the contact angle) is valid here, eq~\eqref{eq:G} can be rewritten as
\begin{equation}
\label{eq:include Young's}
G = \gamma_{LV} A_{LV} + \gamma_{SV}\left(A_{SV} + A_{LS}\right) - \gamma_{LV} A_{LS} \cos\theta - \rho g \Omega z,
\end{equation}
where ($A_{SV} + A_{LS}$) equals solid surface area and thus is a constant.

After an external perturbation, the position of the droplet moves by an infinitesimal distance~$\delta z$ as shown in Figure~\ref{fig:schematic}(B). 
Based on eq~\eqref{eq:include Young's}, the resulting change of total free energy ($\delta G$) is given by
\begin{equation}
\label{eq:delta G}
\delta G = \gamma_{LV} \delta A_{LV} - \gamma_{LV} \delta A_{LS} \cos\theta - \rho g \Omega \delta z,
\end{equation}
where $\delta A_{LV}$ is the change of the liquid-vapor interfacial area and arises from that the wetted fiber occupies part of the apparent volume ($V$) of the droplet, i.e., $\Omega+2\pi r^2 l=V$, where $V=4\pi R^3/3$; 
$\delta A_{LS}$ is the change of the liquid-solid interfacial area due to the movement of three-phase contact points ($P$ and $P'$). 
The energy potential is
\begin{equation}
\label{eq:G potential}
\frac{\delta G}{\delta z} = \gamma_{LV} \frac{\delta A_{LV}}{\delta z} - \gamma_{LV} \frac{\delta A_{LS}}{\delta z} \cos\theta - \rho g \Omega,
\end{equation}
where $A_{LV}=4\pi R^2$, so $\delta A_{LV}/\delta z$ is expressed as
% based on the apparent volume relation $\Omega+2\pi r^2 l=4\pi R^3/3$, as (see ESI section 2 for derivation)
%\zpl{using 0.5 to 2 sentences to explain the core ideas behind eq. (5) and (6). You did an OK job for eq (7) on this regard.}
\begin{equation}
\label{eq:A_LV}
\frac{\delta A_{LV}}{\delta z} \approx 4\pi r^2 \left(\frac{3}{4\pi}\Omega\right)^{-\frac{1}{3}}\frac{\delta l}{\delta z},
\end{equation}
which is based on the apparent volume relation $\Omega+2\pi r^2 l=4\pi R^3/3$ (see ESI section 2 for derivation).
Invoking $A_{LS}=4\pi r l$, $\delta A_{LS}/\delta z$ is
\begin{equation}
\label{eq:A_LS}
\frac{\delta A_{LS}}{\delta z} = 4\pi r\frac{\delta l}{\delta z}.
\end{equation}

$\delta l/\delta z$ in eqs~\eqref{eq:A_LV} and \eqref{eq:A_LS} is a measure of the movement of three-phase contact points when the COM of the droplet is perturbed \cite{zhang2025droplets}. With the spherical assumption, $\delta l/\delta z$ is derived as eq~\eqref{eq:l/s} based on the geometric relationship shown in Figure~\ref{fig:schematic} (see ESI section 3 for detailed derivation):
\begin{equation}
\label{eq:l/s}
\frac{\delta l}{\delta z}=-\frac{\cos \beta}{\cos\left(\beta-\alpha/2\right)}.
\end{equation}
When $\alpha=\pi$, eq~\eqref{eq:l/s} becomes $\delta l/\delta z = -\cot \beta$, recovering the model for $\delta l/\delta z$ developed in \cite{pan2018drop} for a horizontal fiber. Substituting eqs~\eqref{eq:A_LV} and \eqref{eq:A_LS}, along with eq~\eqref{eq:l/s}, into eq~\eqref{eq:G potential} gives
\begin{equation}
\label{eq:G potential2}
\frac{\delta G}{\delta z} \approx 4\pi r \gamma_{LV} \frac{\cos \beta}{\cos\left(\beta-\alpha/2\right)} \left[\cos\theta - r \left(\frac{3}{4\pi}\Omega\right)^{-\frac{1}{3}} \right] - \rho g \Omega.
\end{equation}

An equilibrium condition of the droplet suspended on the fiber is obtained when the total free energy is minimized, and setting $\delta G/\delta z=0$ yields
\begin{equation}
\label{eq:max V}
\Omega \approx 4\pi r \lambda^2 \frac{\cos \beta}{\cos\left(\beta-\alpha/2\right)} \left[\cos\theta - r \left(\frac{3}{4\pi}\Omega\right)^{-\frac{1}{3}}\right],
\end{equation}
where $\lambda =\sqrt{\gamma_{LV}/\left(\rho g\right)}$ is the capillary length of the liquid.
When $\alpha=\pi$, neglecting the the variation of $A_{LV}$ simplifies eq~\eqref{eq:max V} to $\Omega\approx 4\pi r\lambda^2\cos\theta\cot\beta$, which is consistent with the model for~$\Omega$ developed in \cite{pan2018drop} for a horizontal fiber.

Taking $\Omega_\lambda = 4\pi\lambda^3/3$ as characteristic volume of a spherical droplet whose radius is the capillary length, normalizing eq~\eqref{eq:max V} yields the non-dimensional maximum volume
\begin{equation}
\label{eq:norm max V}
\Omega^* = \frac{\Omega}{\Omega_\lambda} \approx 3 r^* \frac{\cos \beta}{\cos\left(\beta-\alpha/2\right)} \left[\cos\theta - \frac{r^*}{{\Omega^*}^{\frac{1}{3}}}\right] = f (\Omega^*),
\end{equation}
where $r^* = r/\lambda$ is non-dimensional fiber radius, and $f(\Omega^*)$ is a function with respect to $\Omega^*$  parameterized by $\beta$, $r^*$, $\theta$, and $\alpha$. 
Once $\beta$ and other parameters are specified, $\Omega^*$ can be obtained by solving eq~\eqref{eq:norm max V}.
$\Omega^*$ in eq~\eqref{eq:norm max V} can also be viewed as a fixed-point for the function $y = f(x)$, where both the independent and dependent variables are the volume of the droplet.
% a function of the volume of the droplet (i.e., $\Omega^* = f(\Omega^*)$), $\Omega^*$ is
% If the right-hand side of eq~\eqref{eq:norm max V} is regarded as a function $f(\Omega^*)$, the equation takes the fixed-point form $\Omega^* = f (\Omega^*)$.
%The solutions of this equation correspond to the fixed points of the mapping, and their stability can be examined by evaluating $|f'(\Omega^*)|$.
In fact, the solution of eq~\eqref{eq:norm max V} provides two fixed points (as exemplified in Cobweb plots, see Figure~S2). 
By comparing the value of $|f'(\Omega^*)|$ with the unity, we find a very small unstable fixed point $\Omega_1^*$, and the larger fixed point $\Omega_2^*$ is stable.
Increasing the size of the droplet from a volume smaller than $\Omega_1^*$ to $\Omega_1^*$ is mathematically unstable, and increasing the size of the droplet beyond $\Omega_2^*$ is impossible, as the volume of the droplet is attracted back to the stable fixed point $x=\Omega_2^*$ even if the size of the droplet is somehow perturbed so that $x > \Omega_2^*$.
Physically, a droplet exceeding $\Omega_2^*$ falls off because gravity overwhelms surface tension.
% or beyond $\Omega_2^*$ is not feasible.
% Droplets with volume below $\Omega_1^*$ are unstable and cannot be sustained on the fiber, whereas droplets exceeding $\Omega_2^*$ fall off because gravity overwhelms surface tension.
Thus, only droplets with volumes within the interval $\Omega_1^*$ and $\Omega_2^*$ can be stably retained on the fiber.
This interpretation is consistent with our experimental procedure: the volume of the droplet is increased using a pipette from an initial volume that is larger than $\Omega_1^*$ until the droplet falls off when its volume approaches $\Omega_2^*$.
% , where droplets can be grown stably toward $\Omega_2^*$. 
A more detailed discussion of this fixed point interpretation is provided in ESI section~4.
%\zpl{double check this equation, why both sides have $\Omega^*$? and eq 9 has the same issue? Can I say that you can actually solve $\Omega^*$ out of this? $\Omega^* = f (\Omega^*)$ can also be viewed as a fixed point for a varying $\Omega^*$. This is very interesting and you should say something about this. also, fixed points can be characterized as stable and unstable. this can be done by analyzing $df/dx$ evaluated at $x = \Omega^*$ and compare it with 1 or -1. Give it a try, and at least put it in the SI, or an appendix. Also look up Cobweb plot. It helps you illustrate the stability of the fixed-points. I think you will have two fixed points, one is very small and it is unstable--- this means that a droplet smaller than that will drop off; The other fixed point has a higher value and it is stable. This is why you can do your experiments: the system allows you to add liquid and increase $x = \Omega^*$ if the droplet is smaller than this fixed point. }

To predict the maximum droplet volume for a given liquid-fiber combination using eq~\eqref{eq:norm max V}, $\beta$ needs to be specified.
Previous studies \cite{lorenceau2004capturing, huang2009equilibrium} often assume $\beta=0^\circ$ (or $90^\circ$ if defined relative to the horizontal) to calculate the theoretical limit of the maximum droplet volume on a horizontal fiber. 
However, Figure~\ref{fig:exp data}(C) demonstrates that $\beta$ varies with $\alpha$ and remains greater than $0^\circ$ even for horizontal fibers.

By analyzing the dependence of $\beta$ on $\alpha$, linear correlations between $\cos\beta$ and $\sin(\alpha/2)$ are observed across all liquid--fiber combinations tested in this work, as shown in the inset of Figure~\ref{fig:exp data}(C). 
Remarkably, the $y$-intercepts of these linear correlations are unity, and their slopes can be well approximated by $2r\lambda^2/\bar{\Omega}$, where $\bar{\Omega}$ is the average maximum droplet volume over the range $\alpha\in[0^\circ,180^\circ]$ for a given liquid--fiber pair, as listed in Table~\ref{tab.A1}. 
%From Figure~\ref{fig:exp data}(A), $\bar{\Omega}$ is around \SI{6}{\micro\liter}, \SI{24}{\micro\liter}, \SI{7}{\micro\liter}, \SI{4}{\micro\liter}, and \SI{2.5}{\micro\liter} for \SI{0.5}{mm} steel-silicone oil, \SI{0.5}{mm} glass-water, \SI{0.33}{mm} PAA-SDS, \SI{0.33}{mm} PAA-silicone oil, and \SI{0.2}{mm} steel-silicone oil, respectively. \zpl{ugly. display them using a table, maybe together with the data in the appendix? } 
Accordingly, a semi-empirical model for $\beta$ as a function of $\alpha$ can be  proposed:
\begin{equation}
\label{eq:semi-empirical beta}
\cos\beta\approx 1-\frac{2r\lambda^2}{\bar{\Omega}}\sin\frac{\alpha}{2}.
\end{equation}
As will be discussed later, $\bar{\Omega}$ does not have to be the exact average maximum volume over the full range of $\alpha$. 
% Any $\Omega$ within $\alpha \in [0^\circ,180^\circ]$ can be taken to represent $\bar{\Omega}$ and still yield a reasonable estimate of $\beta$
% , as $\Omega$ does not vary much.
%\zpl{This last sentence is hard to follow and comes abruptly.}

\subsection{Validation and discussion}
\label{sec: validation}

\begin{figure}[!h]
    \centering
    \includegraphics[width=1\linewidth]{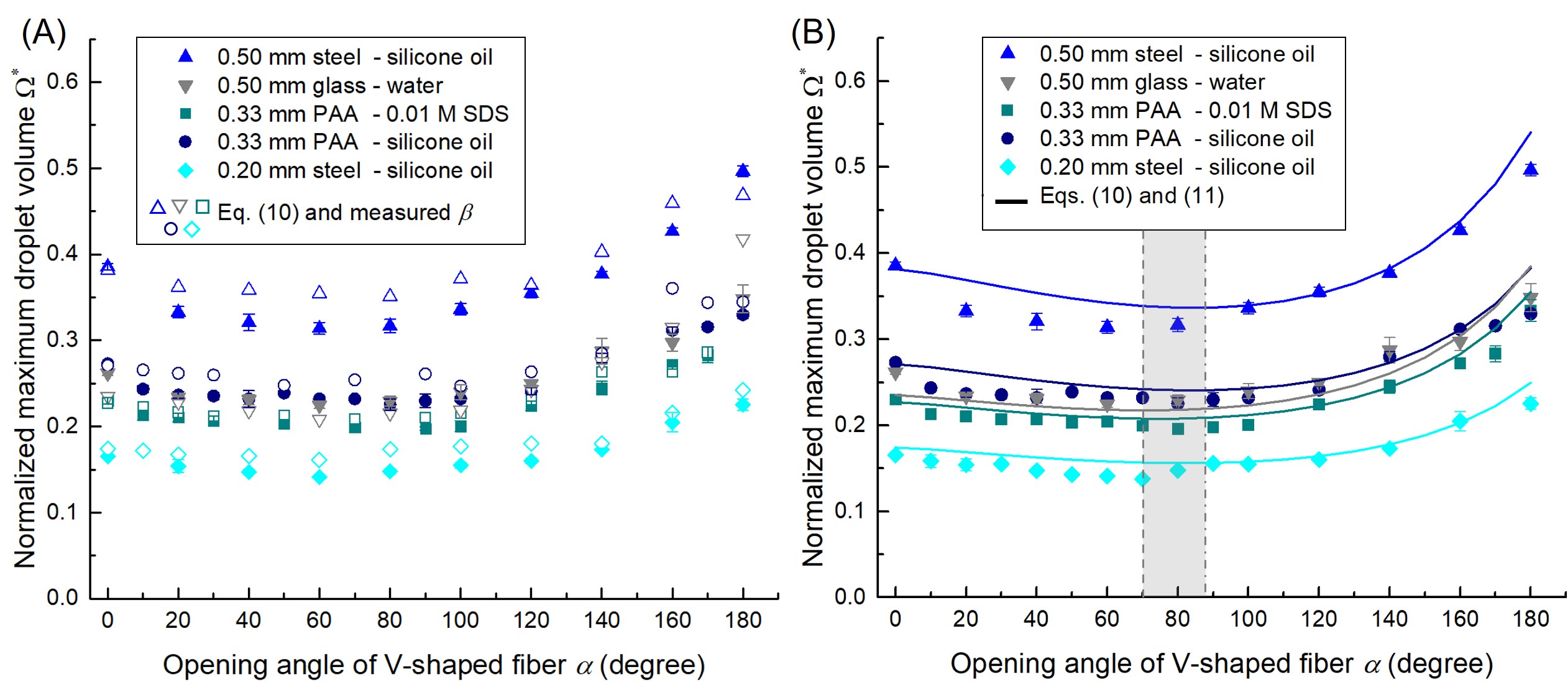}
    \caption{
    %\zpl{I would make the gap between panel A and B a bit wider.} 
    Validation of the model for the normalized maximum droplet volume ($\Omega^*$, eq~\eqref{eq:norm max V}) against the experimental data, which is also normalized by $\Omega_\lambda$. (A) $\Omega^*$ is computed by substituting experimentally measured $\beta$ in Figure~\ref{fig:exp data}(C) into eq~\eqref{eq:norm max V}. (B) $\Omega^*$ is predicted using eq~\eqref{eq:norm max V} with $\beta$ calculated by the semi-empirical model of $\beta$ (eq~\eqref{eq:semi-empirical beta}). The gray-shaded region, bounded by $\alpha=70^\circ$ (the vertical dashed line) and $\alpha=87^\circ$ (the vertical dash-dotted line), indicates the estimated $\alpha$ range for the minimum $\Omega^*$ based on eqs~\eqref{eq:norm max V} and \eqref{eq:semi-empirical beta}.} 
    \label{fig:validation}
\end{figure}

Figure~\ref{fig:validation} validates the developed theoretical model for predicting the maximum droplet volume~($\Omega^*$) against the experimental data in Figure~\ref{fig:exp data}(A).
In Figure~\ref{fig:validation}(A), $\Omega^*$ is calculated, represented by open symbols with the same shape and color scheme as the experimental data, by substituting the measured $\beta$ in Figure~\ref{fig:exp data}(C), along with $r^*$ and $\alpha$, into eq~\eqref{eq:norm max V}.
The model successfully captures the trend of $\Omega^*$ with varying $\alpha$, yielding a mean relative error of 7.4\%.
In Figure~\ref{fig:validation}(B), $\Omega^*$ is predicted (shown as solid lines colored consistently with the experimental legend) using eq~\eqref{eq:norm max V} with $\beta$ from the semi-empirical model (eq~\eqref{eq:semi-empirical beta}).
Remarkably, the use of semi-empirical expression for $\beta$ achieves an even lower mean relative error of 6.1\%.
One possible explanation for this improvement is that direct measurement of $\beta$ from front-view images may be affected by deviations of the droplet shape from the spherical approximation.
Additionally, the critical $\beta$ corresponding to the maximum-volume droplet should act as a dependent intermediate variable, governed by underlying physical mechanisms.
While it remains uncertain whether eq~\eqref{eq:semi-empirical beta} fully captures these mechanisms, it provides a good approximation to the actual droplet behavior.

By analyzing eq~\eqref{eq:semi-empirical beta}, we found that $\beta$ is relatively insensitive to moderate variations in $\bar{\Omega}$ (i.e., the value of $\partial \beta/\partial \Omega$ is moderate).
Even when $\bar{\Omega}$ is replaced with the maximum or minimum value of $\Omega$ within the range $\alpha\in[0^\circ,180^\circ]$, the $\beta$ computed using eq~\eqref{eq:semi-empirical beta} still leads to a reasonable prediction of $\Omega^*$, see Appendix A.
Therefore, the applicability of eq~\eqref{eq:semi-empirical beta} can be extended by using any representative value of $\Omega$ within $\alpha\in[0^\circ,180^\circ]$ for a given liquid-fiber pair in replace of $\bar{\Omega}$.

Based on eqs~\eqref{eq:norm max V} and \eqref{eq:semi-empirical beta}, the dependence of $\Omega^*$ on $\alpha$ can be theoretically predicted. 
Eq~\eqref{eq:norm max V} can be approximated as $\Omega^*\approx 3r^*\cos \beta/\cos\left(\beta-\alpha/2\right)$ for thin fibers ($r^*\ll\left(\Omega^*\right)^{1/3}$) with good liquid wettability ($\cos\theta\approx1$).
Since $r^*$ is constant for a given liquid-fiber pair, the variation of $\Omega^*$ with $\alpha$ is governed by the term $\cos\beta/\cos\left(\beta-\alpha/2\right)$.
In eq~\eqref{eq:semi-empirical beta}, the slope $2r\lambda^2/\bar{\Omega}$ characterizing different liquid--fiber pairs ranges from 0.15 to 0.20 in our experiments.
By substituting the $\beta$ computed from eq~\eqref{eq:semi-empirical beta} with $2r\lambda^2/\bar{\Omega}=0.15\text{--}0.20$ into $\cos \beta/\cos\left(\beta-\alpha/2\right)$, we find that this term initially decreases and then increases as $\alpha$ increases from $0^\circ$ to $180^\circ$, and a minimum is reached at $\alpha=70^\circ$ and $\alpha=87^\circ$ for $2r\lambda^2/\bar{\Omega}=0.15$ and 0.20, respectively (see ESI Section 4).
Accordingly, a gray-shaded region bounded by $\alpha=70^\circ$ and $\alpha=87^\circ$ is presented in Figure~\ref{fig:validation}(B) to indicate the estimated range of $\alpha$ where $\Omega^*$ reaches its minimum.
This estimated range lies within the transition region of $\Omega$ observed in Figure~\ref{fig:exp data}(A).

\section{Conclusion}
\label{sec: Conclusion}
Previous studies have examined the maximum droplet volume that can be sustained on a horizontal fiber (opening angle $\alpha=180^\circ$) \cite{lorenceau2004capturing} and on a $\Lambda$-shaped bend fiber ($\alpha>180^\circ$) \cite{pan2018drop}.
In this work, we address the same problem for V-shaped bent fibers with $\alpha<180^\circ$.
An analytical model based on free energy analysis is developed to predict the maximum droplet volume on V-shaped fibers and is validated using experimental data from five liquid--fiber combinations.
The results show that as $\alpha$ increases from $0^\circ$ to $180^\circ$, the maximum droplet volume follows a non-monotonic trend: it first decreases, then enters a broad transition region around $\alpha \approx 40^\circ \text{--} 100^\circ$, and subsequently increases at larger $\alpha$.

\section*{Appendices}
\appendix
\section{Robustness of eq~(11)}
\label{sec: Appendix}
For a given liquid-fiber pair, the maximum droplet volume $\Omega$ changes with $\alpha$.
Within the range $\alpha \in [0^\circ,180^\circ]$, its minimum and maximum values are represented by $\Omega_{\text{min}}$ and $\Omega_{\text{max}}$, respectively. 
%\zpl{minimum and maximum values of the maximum droplet volume is kind of confusing. why the the maximum volume is a range? I think you may have the same issue in the main text. 0.5 - 1 sentences to explain it.}
Replacing $\bar{\Omega}$ of eq~\eqref{eq:semi-empirical beta} with either $\Omega_{\text{min}}$ or $\Omega_{\text{max}}$ yields
\renewcommand{\theequation}{A\arabic{equation}}
\setcounter{equation}{0}
\begin{equation}
\label{eq:semi-empirical beta1}
\cos\beta\approx 1-\frac{2r\lambda^2}{\Omega_{\text{min}}}\sin\frac{\alpha}{2},
\end{equation}
\begin{equation}
\label{eq:semi-empirical beta2}
\cos\beta\approx 1-\frac{2r\lambda^2}{\Omega_{\text{max}}}\sin\frac{\alpha}{2},
\end{equation}
where $\Omega_{\text{min}}$ and $\Omega_{\text{max}}$ for different liquid--fiber pairs are listed in Table~\ref{tab.A1}.
%, respectively, 
%$4.6\pm0.1\,\si{\micro\liter}$ and $7.3\pm0.1\,\si{\micro\liter}$ for \SI{0.5}{mm} steel-silicone oil,
%$18.7\pm0.3\,\si{\micro\liter}$ and $29.1\pm1.3\,\si{\micro\liter}$ for \SI{0.5}{mm} glass-water, 
%$5.3\pm0.1\,\si{\micro\liter}$ and $9.0\pm0.3\,\si{\micro\liter}$ for \SI{0.33}{mm} PAA-SDS, 
%$3.3\pm0.1\,\si{\micro\liter}$ and $4.8\pm0.1\,\si{\micro\liter}$ for \SI{0.33}{mm} PAA-silicone oil, 
%and $2.0\pm0.1\,\si{\micro\liter}$ and $3.3\pm0.1\,\si{\micro\liter}$ for \SI{0.2}{mm} steel-silicone oil. 
%\zpl{this is very ugly. display them using a table.}

\renewcommand{\thetable}{A\arabic{table}}
\setcounter{table}{0}
\begin{table}[!hb]
\centering
\caption{Minimum, maximum, and mean values of $\Omega$ for different liquid--fiber pairs.}
\label{tab.A1}
\begin{tabular}{c c c c c}
\toprule
Liquid--fiber pair& Fiber diameter [mm]& $\Omega_{\max}$ [$\mu$L]& $\Omega_{\min}$ [$\mu$L]& $\bar{\Omega}$ [$\mu$L]\\
\hline
Silicone oil--steel& 0.50& $7.3 \pm 0.1$& $4.6 \pm 0.1$ & $6.0\pm0.1$\\
Water--glass& 0.50& $29.1 \pm 1.3$& $18.7 \pm 0.3$ & $23.9\pm0.7$\\
SDS--PAA& 0.33& $9.0 \pm 0.3$& $5.3 \pm 0.1$ & $7.2\pm0.2$\\
Silicone oil--PAA& 0.33& $4.8 \pm 0.1$& $3.3 \pm 0.1$ & $4.1\pm0.1$\\
Silicone oil--steel& 0.20& $3.3 \pm 0.1$& $2.0 \pm 0.1$ & $2.7\pm0.1$\\
\bottomrule
\end{tabular}
\end{table}

Figure~\ref{fig:A1} validates the models based on eqs~\eqref{eq:norm max V} and \eqref{eq:semi-empirical beta1} (dash-dotted lines), as well as eqs~\eqref{eq:norm max V} and \eqref{eq:semi-empirical beta2} (dashed lines), against the experimental data.
Replacing $\bar{\Omega}$ in eq~\eqref{eq:semi-empirical beta} with any value of $\Omega$ within the range $\alpha\in[0^\circ,180^\circ]$ yields a model that falls between these two predictions.
Both models successfully capture the variation of $\Omega^*$ with $\alpha$, suggesting that $\bar{\Omega}$ in eq~\eqref{eq:semi-empirical beta} does not need to be strictly defined and can be reasonably approximated by any representative value of $\Omega$ within the range $\alpha\in[0^\circ,180^\circ]$.

\renewcommand{\thefigure}{A\arabic{figure}}
\setcounter{figure}{0}
\begin{figure}[]
    \centering
    \includegraphics[width=1\linewidth]{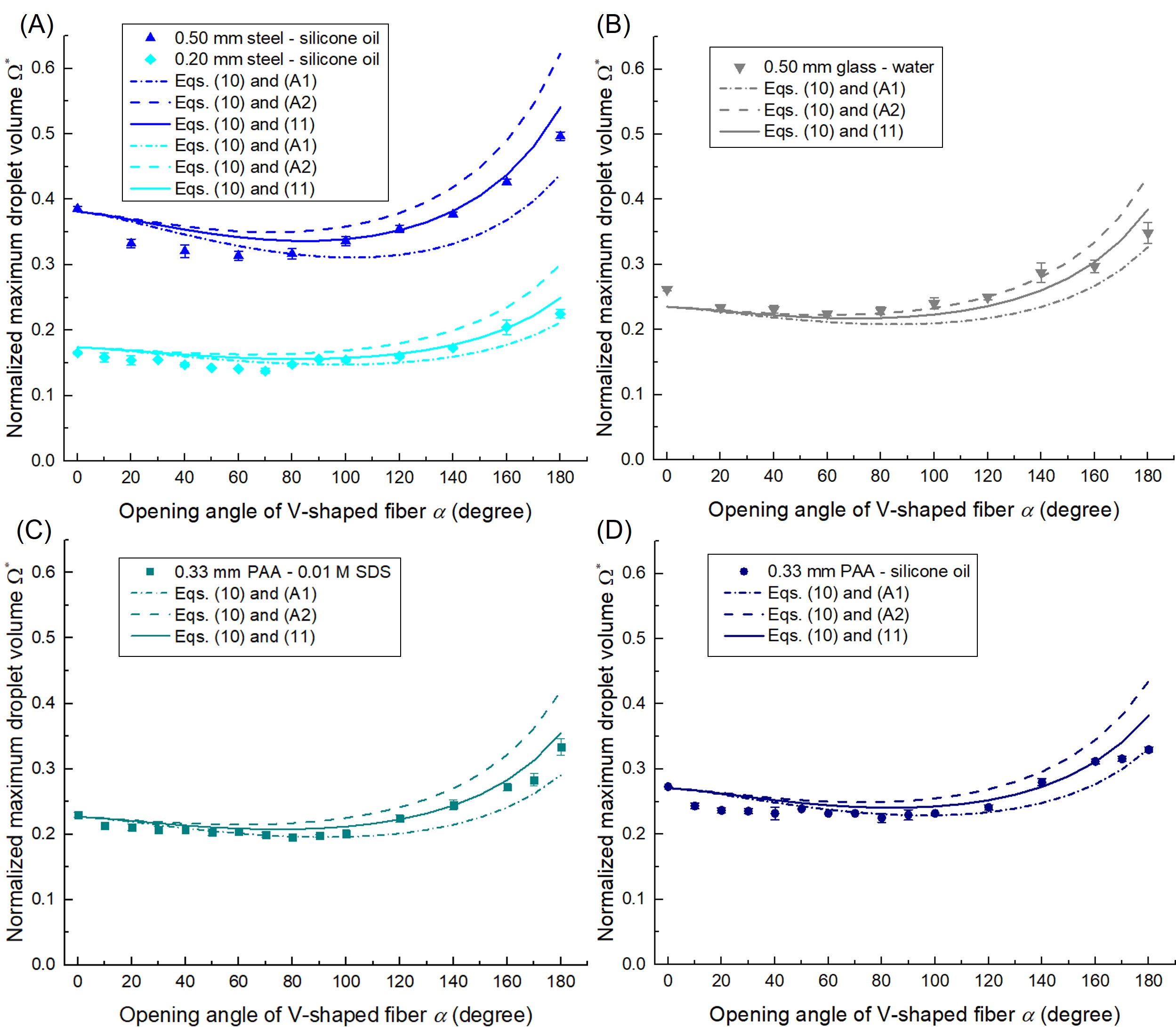}
    \caption{Validation of the models based on eqs~\eqref{eq:norm max V} and \eqref{eq:semi-empirical beta1}, and eqs~\eqref{eq:norm max V} and \eqref{eq:semi-empirical beta2}. (A) Silicone oil on stainless steel fibers with diameters of \SI{0.50}{mm} and \SI{0.20}{mm}. (B) Deionized water on \SI{0.50}{mm} glass fibers. (C) \SI{0.01}{M} SDS solution on \SI{0.33}{mm} PAA-coated fibers. (D) Silicone oil on \SI{0.33}{mm} PAA-coated fibers.}
    \label{fig:A1}
\end{figure}
%\zpl{this section and figure~S2 looks like black magic... very nice tho.  I would put this in the main text as an appendix. }

\section*{Acknowledgement}
The authors would like to express their sincere gratitude for the financial support provided by the Artificial Intelligence for Design Challenge (AI4D) program, funded by the National Research Council Canada. 

\bibliographystyle{unsrt}

\bibliography{reference}

\end{document}